\documentclass[conference,a4paper,onecolumn]{IEEEtran}
\IEEEoverridecommandlockouts

\usepackage{amsmath}
\usepackage{amsthm}
\usepackage{amssymb}
\usepackage{epsfig}
\usepackage{epstopdf}
\usepackage{graphicx}
\usepackage{graphics}
\usepackage{cite}

\newtheorem{thm}{Theorem}
\newtheorem{cor}[thm]{Corollary}

\theoremstyle{definition}

\theoremstyle{remark}

\newtheorem{claim}{\bf Claim}

\newcommand{\snr}{ {\mathsf{snr}} }

\newcommand{\set}[1]{\left\{#1\right\}}

\newcommand\ie{{{i.e.}}}
\newcommand\eg{{{e.g.}}}

\newcommand{\comment}[1]  {}
\def\BE{\begin{equation}}
\def\EE{\end{equation}}
\def\BEA{\begin{eqnarray}}
\def\EEA{\end{eqnarray}}
\newcommand{\pd}[2]{\frac{\partial #1}{\partial #2}}

\newcommand\vn{{\bf n}}

\newcommand\vx{{\bf x}}
\newcommand\vy{{\bf y}}

\newcommand\mA{{\bf A}} 

\newcommand\mI{{\bf I}}

\newcommand{\rmd}{\mathrm{d}}
\newcommand{\rme}{\mathrm{e}}

\ifCLASSINFOpdf
\else
\fi
\begin{document}
%
\title{Low-Density Code-Domain NOMA:\\Better Be Regular
}

\author{\IEEEauthorblockN{Ori Shental
}
\IEEEauthorblockA{Communications Theory Department\\
Bell Labs   \\
Holmdel, New Jersey 07733, USA\\
Email: ori.shental@nokia-bell-labs.com}
\and
\IEEEauthorblockN{Benjamin M. Zaidel
}
\IEEEauthorblockA{Faculty of Engineering\\
Bar-Ilan University\\
Ramat-Gan 52900, Israel\\
Email: benjamin.zaidel@gmail.com}
\and
\IEEEauthorblockN{Shlomo Shamai (Shitz)
}
\IEEEauthorblockA{Department of Electrical Engineering\\
Technion\\ Haifa 32000, Israel\\
Email: sshlomo@ee.technion.ac.il
}}


%


\maketitle

\begin{abstract}
A closed-form analytical expression is derived for the limiting empirical squared singular value density of a spreading (signature) matrix corresponding to sparse low-density code-domain (LDCD) non-orthogonal multiple-access (NOMA) with \emph{regular} random user-resource allocation. The derivation relies on associating the spreading matrix with the adjacency matrix of a large semiregular bipartite graph. For a simple repetition-based sparse spreading scheme, the result directly follows from a rigorous analysis of spectral measures of infinite graphs.  Turning to random (sparse) binary spreading, we harness the cavity method from statistical physics, and show that the limiting spectral density coincides in both cases.
Next, we use this density to compute the normalized input-output mutual information of the underlying vector channel in the large-system limit. The latter may be interpreted as the achievable total throughput per dimension with optimum processing in a corresponding multiple-access channel setting or, alternatively, in a fully-symmetric broadcast channel setting with full decoding capabilities at each receiver.
Surprisingly, the total throughput of regular LDCD-NOMA is found to be not only superior to that achieved with irregular user-resource allocation, but also to the total throughput of \emph{dense} randomly-spread NOMA, for which optimum processing is computationally intractable.
%
In contrast, the superior performance of regular LDCD-NOMA can be potentially achieved with a feasible message-passing algorithm. This observation may advocate employing regular, rather than irregular, LDCD-NOMA in 5G cellular physical layer design.
\end{abstract}


%
\IEEEpeerreviewmaketitle

\section{Introduction}
Non-orthogonal multiple-access (NOMA) is a main enabler of the new radio (NR) design of 5G cellular networks and beyond. The underlying idea is to loosen the paradigm of orthogonal transmissions by allowing different users (or ``layers") to concurrently share the same physical resources, in either time, frequency or space. Consequently, more connections can be supported in massive Machine-Type-Communications (mMTC), or alternatively, a higher total throughput can be achieved in enhanced Mobile Broadband (eMBB) scenarios.

The plethora of NOMA techniques can be roughly categorized into two main classes: signature-domain multiplexing, and power-domain multiplexing (see, \eg, \cite{NOMA:Overview} for a recent overview). In the latter class, signals corresponding to different users are superimposed,
and commonly decoded via successive interference cancellation (SIC). Signature-domain multiplexing is based on distinguishing spreading codes, or interleaver sequences (concatenated with low-rate error-correcting codes).
%

Low-density code-domain (LDCD) NOMA is a prominent sub-category of signature-based multiplexing, which relies on low-density signatures (LDS) \cite{NOMA:LDS}.
Sparse spreading codes comprising a small number of non-zero elements are  employed for linearly modulating each user's symbols over shared physical resources. 
Significant receiver complexity reduction can be achieved by utilizing message-passing algorithms (MPAs) \cite{NOMA:scmaDetection3},
which enable user separation even when the received powers are comparable (as opposed to power-domain NOMA).
Different variants of LDCD-NOMA have recently gained much attention in 5G 3GPP standardization \cite{NOMA:Overview}.
For instance, Sparse-Code Multiple-Access (SCMA)
\cite{NOMA:scmaCodeBook}
further optimizes the low-density sequences to achieve shaping and coding gains by using multidimensional constellations.
In this paper, we chose, however, to focus on the more conventional linear spreading.
Our results can therefore be regarded as providing information-theoretic lower bounds on the expected performance of the various LDCD-NOMA schemes discussed in the literature.

The sparse mapping between users and resources in LDCD-NOMA can be either regular, where each user occupies a fixed number of resources, and each resource is used by a fixed number of users; or irregular, where the respective numbers are random, and only fixed \emph{on average}.
The optimal spectral efficiency of irregular LDCD-NOMA was investigated in\cite{NOMA:YoshidaTanaka}, and shown to reside below the well-known spectral efficiency of dense random-spreading (RS) \cite{NOMA:VerduShamai}. The result stems from
the random nature of the user-resource mapping,
due to which
some users may end up without any designated resources, while some resources may be left unused. This raises the question of whether a regular user-resource mapping leads to any capacity gains. 
Note in this respect that a regular mapping requires some kind of coordination or central scheduling, and may therefore pose some additional practical
challenges.
%

In this paper, we investigate the total achievable throughput per dimension of \emph{regular} LDCD-NOMA, while considering Gaussian signaling, non-fading channels and the large-system limit. We show that it is not only superior to irregular LDCD-NOMA \cite{NOMA:YoshidaTanaka}, but also \emph{outperforms RS-NOMA} \cite{NOMA:VerduShamai} (preliminary observations of similar nature were numerically obtained for binary-inputs in \cite{NOMA:RaymondSaad}).
This result is remarkable, since regular LDCD-NOMA allows for feasible near-optimal (MPA-based) multiuser receiver implementation, whereas the complexity of the optimum receiver for RS-NOMA is prohibitive for large systems. Our information-theoretic observation relies on deriving a closed-form expression for the limiting 
squared singular value density of the underlying channel transfer matrix. For a simple repetition-based sparse spreading scheme, the result directly follows from a rigorous analysis of spectral measures of infinite graphs \cite{NOMA:godsil1988walk}.  For random (sparse) binary spreading, we harness the cavity method from statistical physics \cite{NOMA:PhysRevE.78.031116}, and show that the limiting density coincides in both cases.



The paper is organized as follows. The regular LDCD-NOMA system model is described in Section~\ref{sec_model}. Section~\ref{sec_SPED} derives the limiting
density of the squared singular values of the underlying channel transfer matrix. Section \ref{sec_SPEF}  presents  numerical results for the total achievable throughput.  Finally, Section~\ref{sec_conc} ends this paper with some concluding remarks.

\section{Regular Low-Density Code-Domain NOMA}\label{sec_model}

Consider a system where the signals of $K$ users (``layers") are multiplexed over $N$ shared orthogonal dimensions (resources). Multiplexing is performed while employing randomly chosen sparse spreading signatures of length $N$ (namely, $N$-dimensional vectors). Each sparse signature is assumed to contain a small number of non-zero entries (typically much smaller than $N$), while the remaining entries are set to zero.
%
For the sake of simplicity, but without loss of conceptual scope, we focus henceforth on real-valued channels.
We further restrict the discussion to the following simple generic Gaussian vector channel model, representing the $N$-dimensional received signal at some arbitrary time instance:
\begin{IEEEeqnarray}{rCl}\label{eq_model}
    \vy&=&\frac{1}{\sqrt{d}}\mA\vx+\vn,
\end{IEEEeqnarray}
where $\vx$ is a $K$-dimensional vector comprising the coded symbols of the users. Assuming Gaussian signaling, full symmetry, and no cooperation between encoders corresponding to different users, the input vector $\vx$ is distributed as $\vx\sim\mathcal{N}(\mathbf{0},P_{x}\mI_{K})$, where $P_x$ denotes the power allocated to each user.
The $N\times K$ 
matrix $\mA$ denotes the sparse signature matrix, whose $k$th column represents the spreading signature of user $k$, and its non-zero entries designate the user-resource mapping (namely, user $k$ occupies dimension $n$ if $A_{nk}\neq0$, where $A_{nk}$ denotes the $(n,k)$th entry of $\mA$).
 Finally, $\vn\sim\mathcal{N}(\mathbf{0},\sigma^{2}\mI_{N})$ denotes the $N$-dimensional additive white Gaussian noise (AWGN) vector. The normalization factor of $1/\sqrt{d}$ in \eqref{eq_model}, where $d\in\mathbb{N}^+$, accounts for the sparsity of the signatures (as explained in the sequel). We use $\snr\triangleq{P_x}/{\sigma^2}$ to denote the per-user signal to noise ratio (SNR),
and $\beta\triangleq K/N$ to denote the \emph{system load} (users per dimension).
In the NOMA context we are specifically interested in the overloaded regime of $\beta\geq1$, where user orthogonality is infeasible. However, extension of the results
to the underloaded setting ($\beta<1$) is straightforward.


In LDCD-NOMA, since the signature matrix $\mA$ is sparse, typically only a few of the users' signals collide over any given orthogonal dimension. The regularity assumption dictates that each column of $\mA$ (respectively, row) has \emph{exactly} $d$ (respectively, $\beta d$) non-zero entries.
$\beta$ is chosen here so that $\beta d\in\mathbb{N}^{+}$. We note in this respect that the matrix $\mA$ is dubbed \emph{irregular} when in a corresponding setting the number of non-zero entries in each column (respectively, row) is Poisson distributed with \emph{average} $d$ (respectively, $\beta d$). The non-zero entries of $\mA$ are assumed in this paper to either equal $1$, corresponding to repetition-based spreading; or to be independent identically distributed (i.i.d.) $\set{\pm1}$ random variables, corresponding to random binary spreading.
Thus, the normalization in \eqref{eq_model} ensures that the columns of ${\mA}/{\sqrt{d}}$ have unit norm.
We further note that the signature matrix $\mA$ can be associated with the adjacency matrix of a random bipartite (factor) graph, $\mathcal{A}$, where a user node $k$ and a resource (dimension) node $n$ are connected if and only if $A_{nk}\neq0$.
This factor graph is assumed to be locally tree-like\footnote{
Loosely speaking, this assumption implies that short cycles on the graph are rare. More precisely, this property implies that the random graph converges in law to a random tree
\cite{NOMA:bordenave2010resolvent}.},
similarly to LDPC codes \cite[Chapters 3-4]{NOMA:richardson2008modern}.
Such construction allows for reduced complexity near-optimal multiuser detection, which can be performed iteratively by applying MPAs over the underlying factor graph.

Assuming that the signature matrix $\mA$ is perfectly known at the receiving end, and uniformly chosen randomly and independently per each channel use from the set of $(d,\beta d)$-regular matrices, the performance measure considered in the sequel is the normalized conditional input-output mutual information:
\begin{IEEEeqnarray}{rCl}\label{eq:I}
C_N^{\text{reg}}(\snr,\beta,d) &\triangleq& \frac{1}{N} I(\vx;\vy|\mA) \IEEEnonumber \\ &=& \frac{1}{2N}\, \mathbb{E}_{\mA}\!\set{\log_2\det\left(\mI_N+\frac{\snr}{d}\mA\mA^T\right)}\ .\quad  \label{eq: Definition of the normalized achievable throughput}
\end{IEEEeqnarray}
This quantity corresponds to the total achievable throughput in bits/sec/Hz per dimension with optimum processing if \eqref{eq_model} is interpreted as representing a multiple-access channel (uplink) setting. Alternatively, it may be interpreted as representing the total normalized throughput in a \emph{fully symmetric}  broadcast channel (downlink) setting with superposition coding, where all users are capable of decoding all messages, and perform the same type of joint decoding. We thus conclude that $C_N^{\text{reg}}(\snr,\beta,d)$ is a reasonable choice for characterizing the expected performance of LDCD-NOMA systems, for both the uplink and the downlink.

\section{Spectral Analysis of Regular LDCD-NOMA}\label{sec_SPED}

We now derive a closed-form expression for the limiting empirical eigenvalue distribution of $\mA\mA^{T}/d$ (cf.\ \eqref{eq: Definition of the normalized achievable throughput}), which constitutes the main contribution of this paper. A key step in the derivation is the association
of the signature matrix $\mA$ with a $(d,\beta d)$-semiregular bipartite graph comprising $N$ resource nodes, and $K$ user nodes.
The $(N+K)\times(N+K)$ symmetric adjacency matrix of this bipartite graph can be expressed as
\BE\label{eq_Atilde}
    \tilde{\mA}\triangleq\begin{pmatrix}
                \mathbf{0} & \mA \\
                \mA^{T} & \mathbf{0} \\
              \end{pmatrix},
\EE
while
\BE
    \tilde{\mA}^2 = \begin{pmatrix}
                \mA\mA^{T} & \mathbf{0} \\
                \mathbf{0} & \mA^{T}\mA\\
              \end{pmatrix}. \label{eq: tilde-A-sq}
\EE
Thus, observing that the eigenvalues of $\tilde{\mA}^2$ are the eigenvalues of $\mA\mA^{T}$ together with those of $\mA^{T}\mA$ (and for $K>N$ the latter are the same as those of $\mA\mA^{T}$ with additional $K-N$ zeros), the spectral density of $\mA\mA^{T}$ (and thus of $\mA\mA^{T}/d$) can be easily derived from the spectral density of $\tilde{\mA}^2$ (see, \eg, \cite{NOMA:Far}). We now proceed with the statement of the main results.

Consider first repetition-based spreading, for which $\mA$ is a sparse $(0,1)$-matrix.

\begin{thm}\label{thm_main1}
Let $\mA$ be a random $N\times K$ sparse $(0,1)$-matrix with
exactly $d$ $1$'s in each column, exactly $\beta d$ $1$'s in each row (where $d,\beta d\in\mathbb{N}^+$, and $1<d, \beta d <\infty$),
and a locally tree-like
associated $(d,\beta d)$-semiregular bipartite graph $\mathcal{A}$ (cf.\ \eqref{eq_Atilde}).
Then, for $K\ge N$ and $N,K\to\infty$ with $K/N=\beta <\infty$, the empirical distribution of the eigenvalues of $\mA\mA^{T}/d$ converges weakly to a distribution whose density function is given by
%
\BE\label{eq_pdf_start}
\rho(\lambda,\beta,d)= \begin{cases}
    \frac{\beta}{2\pi}\frac{\sqrt{\big(d\tau-(\xi-1)^2\big)\big((\xi+1)^2-d\tau\big)}}{\tau\lambda} \, , & \lambda^{-}\le \lambda\le\lambda^{+},\\ 
    0\, , & \text{otherwise}\, ,
  \end{cases}
\EE
where $\lambda^{\pm}=(\sqrt{\alpha}\pm\sqrt{\gamma})^2$, $\alpha\triangleq\frac{d-1}{d}$, $\gamma\triangleq\frac{\beta d-1}{d}$, $\tau\triangleq \beta d-\lambda$, and $\xi\triangleq d\sqrt{\alpha\gamma}$.
\end{thm}
\begin{IEEEproof}[Proof Outline]
The result relies on the properties
of the associated bipartite graph of $\mA$.
More specifically, Godsil and Mohar \emph{rigorously} derive in \cite[Corollary 4.5]{NOMA:godsil1988walk} the limiting eigenvalue distribution of the adjacency matrix of large semiregular bipartite graphs, which asymptotically weakly converge to a bipartite Galton-Watson tree \cite{NOMA:bordenave2010resolvent} (see also \cite{NOMA:mizuno2003semicircle} for an alternative proof).
The result gives the limiting spectral density of $\tilde{\mA}$ in \eqref{eq_Atilde}, from which
the spectral density of $\tilde{\mA}^2$ in \eqref{eq: tilde-A-sq} is readily obtained. Finally, the density \eqref{eq_pdf_start} is derived via the relation between the eigenvalues of $\tilde{\mA}^2$ and $\mA\mA^{T}$, as discussed at the beginning of this section.
\end{IEEEproof}

Next, we turn to random (sparse) binary spreading. For this setting,
the existence of a limiting distribution follows from \cite{NOMA:bordenave2010resolvent}.
The derivation of the spectral density relies here on the heuristic cavity method from statistical physics, as initially developed by M\'{e}zard \emph{et al.}~\cite{NOMA:mezard1986sk}. At the heart of the cavity approach
is the relation between the sought after spectral density, and an appropriately defined partition function (as clarified in the sequel).
When the graph associated with $\mA$ is locally tree-like (essentially, with no short cycles going through a typical node),  if one removes a node from the graph (thus generating a ``cavity"), the joint distribution of its neighborhood factorizes (subject to the associated Boltzmann distribution).
This produces a set of recursive equations, out of which the cavity distributions can be obtained, and
the method is expected to be asymptotically exact for tree-like graphs\cite{NOMA:PhysRevE.78.031116}.
\emph{Postulating} the validity of the cavity method for our setting, we obtain the following result (stated as a ``claim").
\begin{claim}\label{thm_main2}
Let $\mA$ be defined as in Theorem~\ref{thm_main1}, but with $\pm1$ non-zero entries (rather than $1$'s).
Then, for $K\ge N$ and $N,K\to\infty$ with $K/N=\beta <\infty$, the empirical distribution of the eigenvalues of $\mA\mA^{T}/d$ converges weakly to a distribution whose density function is
identical to that
given in \eqref{eq_pdf_start}.
\end{claim}
\begin{IEEEproof}[Proof Outline]
The spectral density of $\tilde{\mA}$ in \eqref{eq_Atilde}, \ie,
\BE
    \rho(\lambda;\tilde{\mA})\triangleq\frac{1}{N+K}\sum_{i=1}^{N+K}\delta(\lambda-\lambda_{i}^{\tilde{\mA}}) \ ,
\EE
where $\{\lambda_{i}^{\tilde{\mA}}\}$ are the eigenvalues of $\tilde{\mA}$, can be expressed via the Edwards-Jones identity \cite{NOMA:PhysRevE.78.031116} as
\BE\label{eq_Edwards}
    \rho(\lambda;\tilde{\mA})=\frac{2}{\pi}\lim_{\epsilon\to0^{+}}\frac{1}{N+K}\,\mathrm{Im}{\left.\left[\pd{}{z}\ln\mathcal{Z}(z;\tilde{\mA})\right]\right|_{z=\lambda+j\epsilon}}.
\EE
Here, the partition function $\mathcal{Z}(z;\tilde{\mA})$ is defined via Gaussian integrals
as (see \cite{NOMA:PhysRevE.78.031116} and references therein)
\BE
    \mathcal{Z}(z;\tilde{\mA})
    =\int\Bigg[\prod_{i=1}^{N+K}\frac{\rmd \tilde{x}_{i}}{\sqrt{2\pi}}\, \rme^{-\mathcal{H}(\tilde{\vx};z,\tilde{\mA})}\Bigg] \ ,
    \label{eq: Definition of partition function}
\EE
with the Hamiltonian
\BE
    \mathcal{H}(\tilde{\vx};z,\tilde{\mA})
    =\frac{z}{2}\sum_{i=1}^{N+K}\tilde{x}_{i}^{2}-\frac{1}{2}\sum_{i\neq j}
\tilde{A}_{ij}\tilde{x}_{i}\tilde{x}_{j} \ .
\EE
Ignoring convergence aspects in \eqref{eq: Definition of partition function} as in \cite{NOMA:PhysRevE.78.031116,NOMA:mezard1986sk}, the spectral density can then be expressed as 
\BE\label{eq_p}
    \rho(\lambda;\tilde{\mA})=-\frac{1}{\pi}\lim_{\epsilon\to0^{+}}\frac{1}{N+K}\sum_{i=1}^{N+K}\mathrm{Im}\left.\left[ \langle {\tilde{x}_{i}^{2}}\rangle_{z}\right]\right|_{z=\lambda+j\epsilon},
\EE
where $\langle\cdot\rangle_{z}$ denotes averaging over the Boltzmann distribution
$\exp(-\mathcal{H}(\tilde{\vx};z,\tilde{\mA}))/\mathcal{Z}(z;\tilde{\mA})$.
The next step at this point is to employ the cavity method, which provides an approximate approach for computing the variances
$\set{\langle {\tilde{x}_{i}^{2}}\rangle_{z}}$
in \eqref{eq_p}, and is expected, as said, to be exact in the large system limit when applied to tree-like graphs.

Omitting details for conciseness, we note that the cavity analysis produces a set of self-consistent equations given in terms of the ``cavity variances'' $\{\Delta_{n}^{(k)}(z)\}$ \cite{NOMA:PhysRevE.78.031116}:
\BE
    \Delta_{n}^{(k)}(z)=\frac{1}{z-\sum_{j\in\partial n\backslash k}\tilde{A}_{nj}^2\Delta_{j}^{(n)}(z)} \ , \ k\in\partial n,
\EE
where $\partial n$ is the set of neighbors of node $n$, $n=1,\dots,N+K $, and $\partial n\backslash k$ denotes the set $\partial n$ with the node $k$ removed. The variances in \eqref{eq_p} then read for $n=1,\dots,N+K$ \cite{NOMA:PhysRevE.78.031116}
\BE
\langle {\tilde{x}_{n}^{2}}\rangle_{z}=
    \tilde{\Delta}_{n}(z)\triangleq\frac{1}{z-\sum_{j\in\partial n}\tilde{A}_{nj}^2\Delta_{j}^{(n)}(z)} \ ,
\EE
from which the spectral density can be derived. Note that the cavity analysis implies that the node variances (and hence \emph{the spectral density} of $\tilde{\mA}$)  depend only on the \emph{squared values} of the non-zero entries of $\tilde{\mA}$.

Recall at this point that we are, in fact, interested in the spectral density of $\mA\mA^{T}$, rather than $\tilde{\mA}$. 
Identifying the limiting mean variance in \eqref{eq_p} as the Cauchy transform  $G_{\tilde{\mA}}(z)$  of the asymptotic eigenvalue distribution of $\tilde{\mA}$ \cite{NOMA:Far}, we now invoke the following relation between $\tilde{\mA}$ and $\tilde{\mA}^2$ (cf.\ \eqref{eq: tilde-A-sq}):
\BE
    G_{\tilde{\mA}}(z)=zG_{\tilde{\mA}^{2}}(z^2) \ . \label{eq: Relation between Cauchy transforms}
\EE
Using \eqref{eq: Relation between Cauchy transforms},
the relation between the eigenvalues of $\tilde{\mA}^2$ and those of $\mA\mA^{T}/d$,
and relying on the full symmetry between the ``cavity variances" corresponding to each type of nodes (namely, either resource nodes or user nodes), one eventually obtains the following simplified set of cavity equations:
\begin{IEEEeqnarray}{rCl}
    \Delta(z) &=& \frac{1}{z-\frac{\gamma}{1-\alpha\Delta(z)}} \ , \\
    \tilde{\Delta}(z)&=&\frac{1}{z-\frac{\beta}{1-\alpha\Delta(z)}} \ ,
\end{IEEEeqnarray}
where $\alpha$ and $\gamma$ are as in \eqref{eq_pdf_start}. Here $\tilde{\Delta}(z)$ represents the Cauchy transform of the limiting eigenvalue distribution of $\mA\mA^{T}/d$.
%
Solving the above equations, and applying the Stiletjes inversion formula \eqref{eq_p} on  $\tilde{\Delta}(z)$, we finally get \eqref{eq_pdf_start}.
This concludes the proof of
Claim \ref{thm_main2}.
\end{IEEEproof}

The cavity method is a powerful computational tool, yet heuristic in nature (similarly to the ``replica method" \cite{NOMA:mezard1986sk}). We note in this respect that the fact that the cavity-based result of
Claim
\ref{thm_main2} coincides with that of Theorem \ref{thm_main1}, which is based on \emph{fully rigorous arguments}, supports the validity of the cavity method in the current setting.
%
%
To further validate our results, we examine two special cases of interest.

Let the matrix $\mA$ be defined as in Theorem~\ref{thm_main1} with $\beta=1$. Consequently, we have $\alpha=\gamma=(d-1)/d$.
Then, the limiting spectral density in \eqref{eq_pdf_start} reads
\BE\label{eq_KM}
\rho(\lambda,1,d)= \begin{cases}
    \frac{d\sqrt{4(d-1)-d\lambda}}{2\pi(d-\lambda)\sqrt{d\lambda}} \, , & 0\le \lambda\le\frac{4(d-1)}{d} \, ,\\
    0 \, , & \text{otherwise},
  \end{cases}
\EE
which is exactly the probability density function of the squared singular values of the well-known Kesten-McKay law \cite{NOMA:MCKAY1981203} for regular square $(0,1)$-matrices.


As a second special case, let the matrix $\mA$ be defined either as in Theorem~\ref{thm_main1} or in
Claim
\ref{thm_main2}. Then, as $d\to\infty$, the limiting spectral density \eqref{eq_pdf_start} converges to
\BE
\rho(\lambda,\beta,d\to\infty)= \begin{cases}
    \frac{\sqrt{(\lambda-\lambda^{-})(\lambda^{+}-\lambda)}}{2\pi\beta\lambda}, & \lambda^{-}\le \lambda\le\lambda^{+} \, , \\
    0, & \text{otherwise}\, ,
  \end{cases}
\EE
where $\lambda^{\pm}=(1\pm\sqrt{\beta})^2$.
As expected \cite{NOMA:PhysRevE.78.031116}, this density is nothing but the Mar\u{c}enko-Pastur law \cite{NOMA:tulino2004random}. We note that similar observations
were made in \cite{NOMA:dumitriu2014marvcenko}, while considering $(0,1)$ matrices (as in Theorem \ref{thm_main1}), and focusing on the case in which $d\to\infty$ while $d/N\to 0$ at an appropriate rate (see therein).

\comment{\subsubsection{1D-Wyner Model with Maximal Interference}
\begin{cor}
TBA (trick: via removing diagonal terms from matrix $\mA$)
\end{cor}
}

\begin{figure}[!htb]
\centering
    \includegraphics[scale=0.55]{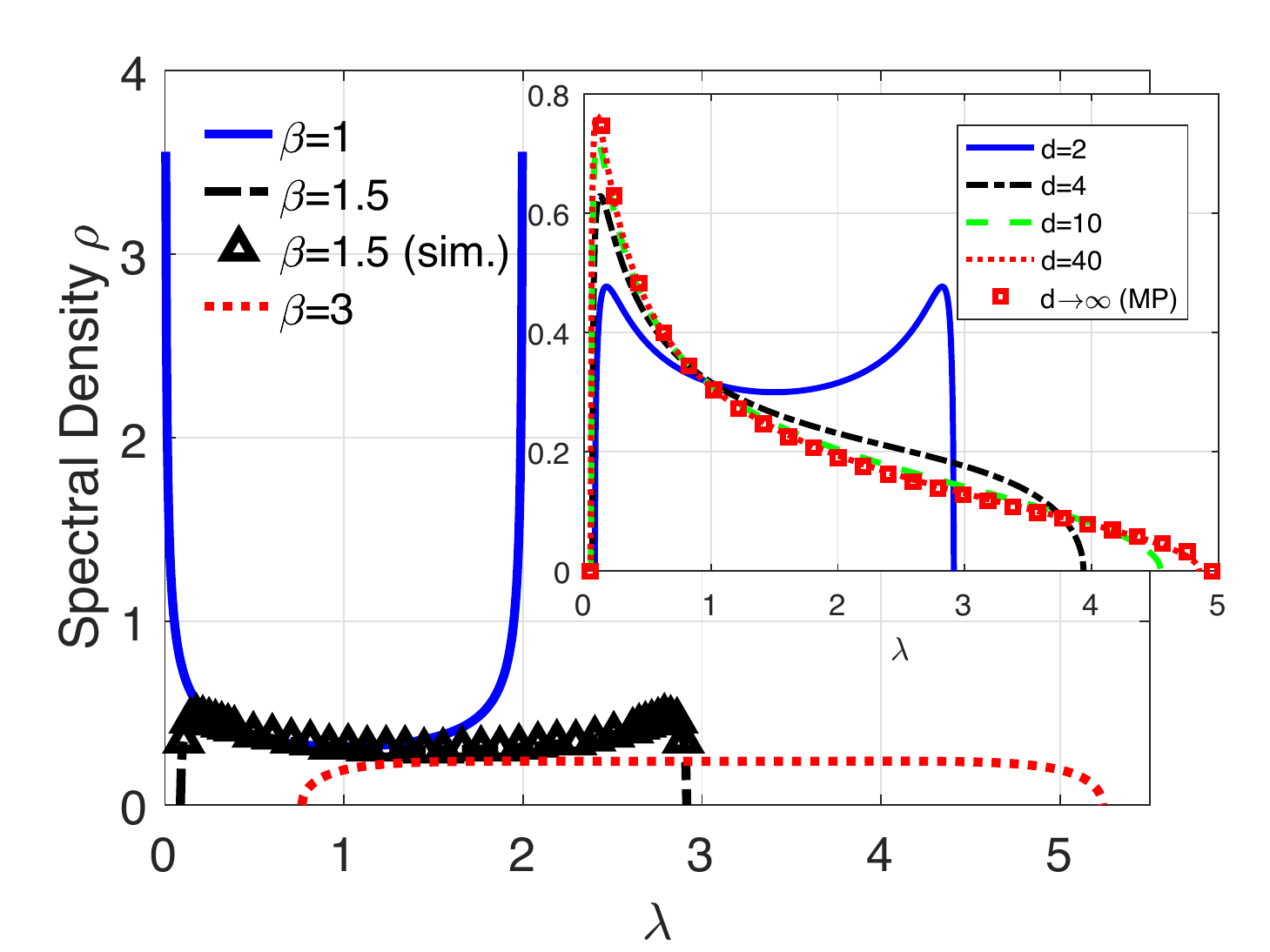}
  \caption{Spectral density of $\mA\mA^{T}/d$  \eqref{eq_pdf_start} for $d=2$ and $\beta=1,1.5,3$ (simulation results are included for $\beta=1.5$).
 Inset: Limiting spectral density of $\mA\mA^{T}/d$ \eqref{eq_pdf_start} for $\beta=1.5$ and sparsity constants $d=2,4,10,40$.
The Mar\u{c}enko-Pastur density ($d\to\infty$) is also shown for comparison.}\label{fig_sped1}
\end{figure}


To get more insight on our limiting results, Fig.\ \ref{fig_sped1} and its inset illustrate the spectral density for several pairs of load $\beta$ and sparsity $d$.
Fig.\ \ref{fig_sped1} shows the limiting spectral density for $d=2$, and loads $\beta=1$ (the Kesten-McKay law), $\beta=1.5$, and $\beta=3$.
For $\beta=1.5$, the analytical result \eqref{eq_pdf_start} is also compared  to the average empirical density obtained via Monte Carlo simulations, while considering a regular sparse spreading matrix of dimensions $2600\times3900$, and $1000$ matrix realizations (as anticipated, simulation results for either $(0,1)$ entries, or $(0,\pm1)$ entries, yield indistinguishable curves). An excellent match to the analytical results is observed.
Trivial eigenvalues that appear in finite-dimensional simulations, but have vanishing weight in the large-system limit, were discarded from the figure (\eg, the maximal eigenvalue of $\mA\mA^T/d$ which equals $\beta d$).
The inset of Fig.\ \ref{fig_sped1} demonstrates the convergence of the limiting spectral density to the Mar\u{c}enko-Pastur law for increasing values of $d$ and $\beta=1.5$.
Note that setting $d=40$ already seems large enough for the limiting density \eqref{eq_pdf_start}
to approach the Mar\u{c}enko-Pastur law, which holds for
dense spreading matrices (corresponding to RS-NOMA), where the average number of non-zero entries in each row, or column, is \emph{proportional} to the respective dimension
\cite{NOMA:VerduShamai} (a similar observation was reported for the irregular setting in \cite{NOMA:YoshidaTanaka}).

\section{Numerical Results}\label{sec_SPEF}

\begin{figure}[!htb]
\centering
    \includegraphics[scale=0.5]{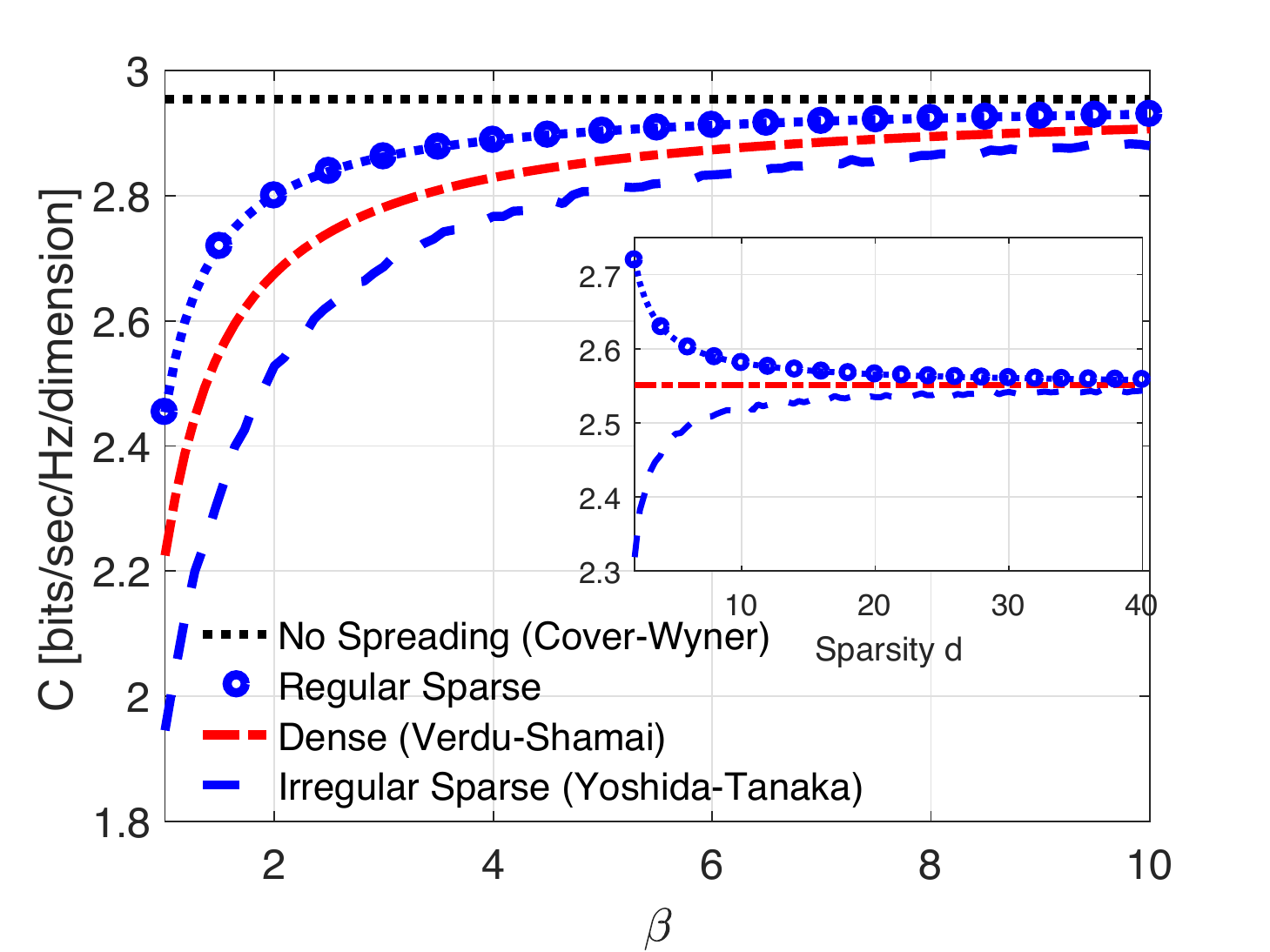}

  \caption{Limiting normalized total throughput vs.\ the system load $\beta$ for $\frac{E_{b}}{N_{0}}=10$dB. The results for either regular or irregular LDCD-NOMA are plotted for $d=2$.
Inset: Limiting normalized total throughput vs.\ the sparsity parameter $d$ for $\frac{E_{b}}{N_{0}}=10$dB and $\beta=1.5$.}\label{fig_spef1}
\end{figure}

\begin{figure}[!htb]
\centering
    \includegraphics[scale=0.5]{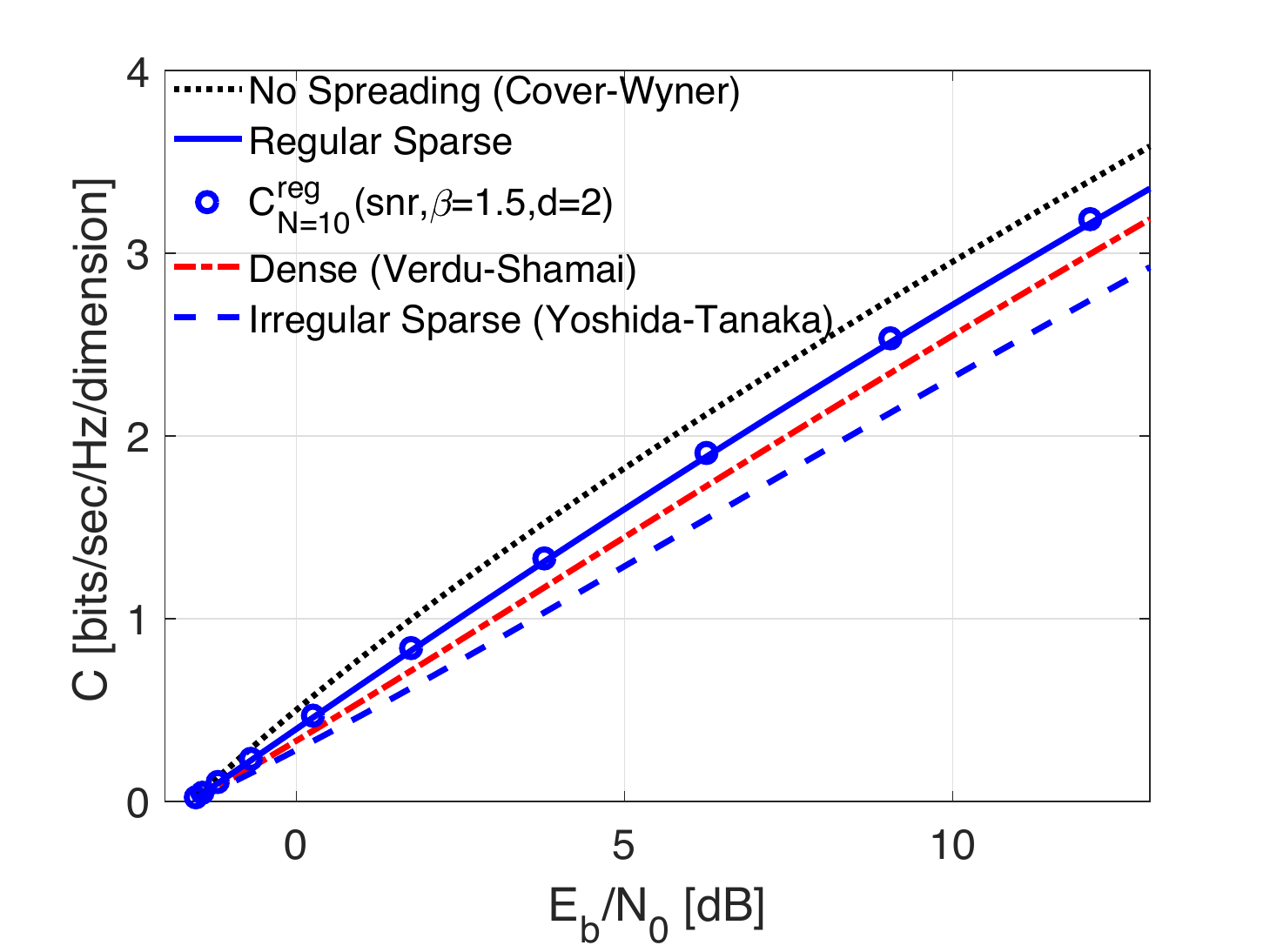}
  \caption{Limiting normalized total throughput vs.\ $\frac{E_{b}}{N_{0}}$ for $\beta=1.5$ and $d=2$.}\label{fig_spef2}
\end{figure}

With the limiting eigenvalue distribution results of Theorem \ref{thm_main1} and
Claim
\ref{thm_main2}, the limiting normalized total achievable throughput of regular LDCD-NOMA  in \eqref{eq: Definition of the normalized achievable throughput} can be analytically computed as (see, e.g., \cite[Proposition III.2]{NOMA:VerduShamai})
%
\begin{IEEEeqnarray}{rCl}
C^{\text{reg}}(\snr,\beta,d)&\triangleq&\lim_{N\to\infty}C_N^{\text{reg}}(\snr,\beta,d) \IEEEnonumber \\&=&\frac{1}{2}\int_{\lambda^{-}}^{\lambda^{+}}\log_2(1+\snr\lambda)\rho(\lambda,\beta,d) \, \rmd\lambda \ , \quad \label{eq: Limiting total throughput}
\end{IEEEeqnarray}
where $\rho(\lambda,\beta,d)$ is given in \eqref{eq_pdf_start}.
Fig.\ \ref{fig_spef1}, its inset, and Fig.\ \ref{fig_spef2} show $C^{\text{reg}}(\snr,\beta,d)$  {(computed via numerical integration)}, plotted, respectively, as a function of the load $\beta$, the sparsity parameter $d$, and $\frac{E_b}{N_0}$ (via the relation $\frac{E_b}{N_0}=\beta\snr/(2C^\text{reg}(\snr))$, see \cite{NOMA:VerduShamai}). We  note in this respect that the regularity constraints imply that only values for which $\beta d \in \mathbb{N}^+$ and $\beta d>1$ have a practical meaning.  The latter are marked by circles in Fig.\ \ref{fig_spef1}.
%
For the sake of comparison we also included in the figures the spectral efficiencies of irregular LDCD-NOMA \cite[Eqs.\ (28)--(32)]{NOMA:YoshidaTanaka}, overloaded dense (\ie, $d\to\infty$) RS-NOMA \cite[Eq.\ (9)]{NOMA:VerduShamai}, and the Cover-Wyner upper bound \cite[Eq.\ (4)]{NOMA:VerduShamai}, which corresponds to the absence of spreading.

The results indicate that regular LDCD-NOMA not only outperforms its corresponding irregular LDCD-NOMA scheme, but even outperforms dense RS-NOMA \cite{NOMA:VerduShamai}. More specifically, Figs.\ \ref{fig_spef1} and \ref{fig_spef2} show that regular LDCD-NOMA exhibits a throughput gain of about 20\% over irregular LDCD-NOMA,
at a typical working point of $\beta=1.5$, $\frac{E_{b}}{N_{0}}=10$dB and $d=2$. For $\beta=1$, a throughput gain close to $30\%$  is observed. We further note that the inset of Fig.\ \ref{fig_spef1} shows how the limiting throughput of regular LDCD-NOMA approaches  that of dense RS-NOMA from above, as the sparsity parameter $d$ increases. Irregular LDCD-NOMA exhibits a similar behavior but from below.
Fig.\ \ref{fig_spef2} also shows simulation results for $C_N^{\textrm{reg}}(\snr,\beta,d)$ of \eqref{eq: Definition of the normalized achievable throughput} with $N=10$ (averaged over $10000$ matrix realizations with binary sparse spreading). The excellent match with the analytical results
establishes the relevance of our asymptotic analysis for practical system dimensions (e.g., \cite{NOMA:scmaCodeBook}).

\section{Concluding Remarks}\label{sec_conc}

Understanding the fundamental information-theoretic performance limits of the various NOMA technologies suggested in recent years, is \emph{crucial} for efficient state-of-the-art designs of future 5G and beyond cellular communication systems.
The topic is therefore timely, and of both theoretical and practical interest.
An insightful attempt in this framework was presented in this paper, by examining the advantages of regular LDCD-NOMA transmission schemes. Harnessing tools from the spectral theory of large random graphs, as well as the heuristic, yet powerful, cavity  method of statistical physics, the achievable total throughput of regular LDCD-NOMA was \emph{analytically} investigated in the large system limit. The results demonstrate that regular sparse spreading may potentially lead to significant performance enhancement over conventional irregular sparse spreading (and in fact also over legacy dense RS-NOMA). Extensions of the results to account for non-binary spreading, fading channel models, and suboptimal receiver designs, are currently investigated.

\section*{Acknowledgment}
The work of B.\ M.\ Zaidel and S.\ Shamai (Shitz) was supported 
by the Heron consortium via the Israel Ministry of Economy
and Industry.
O.\ Shental would like to thank 
R.\ Valenzuela, D.\ Ezri, and E.\ Melzer for useful discussions.





%


\end{document}